# How to use Big Data technologies to optimize operations in Upstream Petroleum Industry

**Abdelkader Baaziz**
Cadre Supérieur at Sonatrach
Ph.D. Student and Resercher at IRSIC Laboratory, Aix-Marseille University
E-mail: abdelkader.baaziz@etu.univ-amu.fr

**Luc Quoniam**
Professor at University of Sud Toulon-Var
E-mail: mail@quoniam.info

## Abstract

"Big Data is the oil of the new economy" is the most famous citation during the three last years. It has even been adopted by the World Economic Forum in 2011. In fact, Big Data is like crude! It's valuable, but if unrefined it cannot be used. It must be broken down, analyzed for it to have value.

But what about Big Data generated by the Petroleum Industry and particularly its upstream segment?

Upstream is no stranger to Big Data. Understanding and leveraging data in the upstream segment enables firms to remain competitive throughout planning, exploration, delineation, and field development.

Oil & Gas Companies conduct advanced geophysics modeling and simulation to support operations where 2D, 3D & 4D Seismic generate significant data during exploration phases. They closely monitor the performance of their operational assets. To do this, they use tens of thousands of data-collecting sensors in subsurface wells and surface facilities to provide continuous and real-time monitoring of assets and environmental conditions. Unfortunately, this information comes in various and increasingly complex forms, making it a challenge to collect, interpret, and leverage the disparate data. As an example, Chevron's internal IT traffic alone exceeds 1.5 terabytes a day.

Big Data technologies integrate common and disparate data sets to deliver the right information at the appropriate time to the correct decision-maker. These capabilities help firms act on large volumes of data, transforming decision-making from reactive to proactive and optimizing all phases of exploration, development and production. Furthermore, Big Data offers multiple opportunities to ensure safer, more responsible operations. Another invaluable effect of that would be shared learning.

The aim of this paper is to explain how to use Big Data technologies to optimize operations. How can Big Data help experts to decision-making leading the desired outcomes?

# How to use Big Data technologies to optimize operations in Upstream Petroleum Industry

## I. Introduction

The Oil & Gas industry is extremely competitive and highly regulated environment. Against this uncertain environment characterized by the eternal necessity to renewal reserves of natural resources, fluctuating demand and price volatility, Oil & Gas companies need to increase production, optimize costs and reduce the impact of environmental risks.

Oil & Gas upstream sector is complex, data-driven business with data volumes growing exponentially (Feblowitz, 2012). Upstream organizations work simultaneously with both structured and unstructured data. They must capture and manage more data than ever and are struggling to store, analyze and get useful information from these huge volumes of data. Under these conditions, the traditional analysis tools would fail but with the appropriate infrastructure and tools, Oil & Gas companies can get measurable value from these data.

The aim of this paper is to show how Big Data can be used to gain valuable operational insight in the Oil & Gas industry and to assist in decision-making in different activities of its upstream sector.

## II. Big Data Definition

The first definition of Big Data was developed by Meta Group (now part of Gartner) by describing their three characteristics called "3V": Volume, Velocity and Variety. Based on data quality, IBM has added a fourth V called: Veracity. However, Oracle has added a fourth V called: Value, highlighting the added value of Big Data (Baaziz & Quoniam, 2013a).

Big Data is defined by six characteristics called "6V": Volume, Velocity, Variety, Variability Veracity and Value.

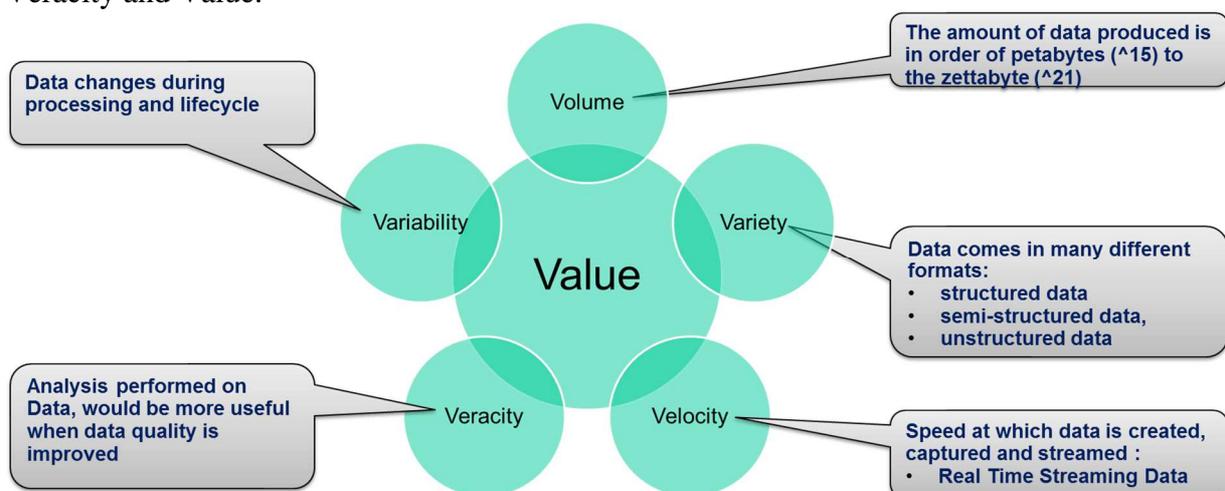

Fig. 1. The "6V" of Big Data

*"Big Data technologies describe a new generation of technologies and architectures, designed to economically extract value from very large volumes of a wide variety of data, by enabling*

*high velocity capture, discovery and/or analysis (Feblowitz, 2012), while ensuring their veracity by an automatic quality control in order to obtain a big value".*

These technologies are essentially based on the Apache™ Hadoop® project that's open-source software for reliable, scalable, distributed computing (Hadoop, 2013).

The Apache Hadoop software library is a framework that allows for the distributed processing of large data sets across clusters of computers using simple programming models. It is designed to scale up from single servers to thousands of machines, each offering local computation and storage. Rather than rely on hardware to deliver high-availability, the library itself is designed to detect and handle failures at the application layer, so delivering a highly-available service on top of a cluster of computers, each of which may be prone to failures (Hadoop, 2013).

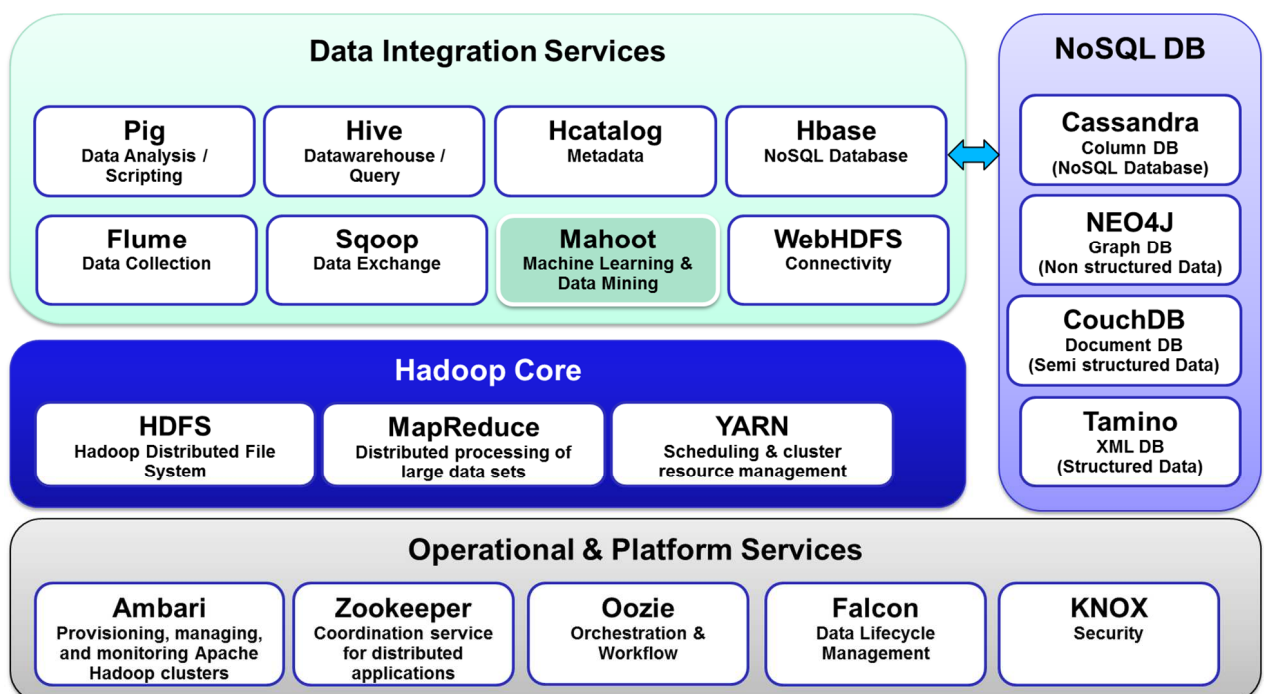

Fig. 2. Big Data technologies are based on Hadoop Ecosystem (Hadoop is an Open source project)

### III.    Big Data in the Oil & Gas Upstream Sector

#### 1.   Overview

Upstream is no stranger to Big Data. Oil & Gas companies use thousands of sensors installed in subsurface wells and surface facilities to provide continuous data-collecting, real-time monitoring of assets and environmental conditions (Brulé, 2013).

The data volume is coming from sensors, spatial and GPS coordinates, weather services, seismic data, and various measuring devices. "Structured" data is handled with specific applications used to manage surveying, processing and imaging, exploration planning, reservoir modeling, production, and other upstream activities. But much of this data is "unstructured" or "semi-structured" such as emails, word processing documents, spreadsheets, images, voice recordings, multimedia, and data market feeds, which means it's difficult or costly to either store in traditional data warehouses or routinely query and analyze. In this case, appropriate tools for Big Data should be used (Hems & al., 2013).

To support the real-time decision-making, Oil & Gas companies need tools that integrate and synthesize diverse data sources into a unified whole. Being able to process Big Data makes it possible to derive insight from the relationships that will surface when all of these sources are processed as a whole. But to unlock this value, Oil & Gas companies need access to the appropriate technology, tools, and expertise (Baaziz & Quoniam, 2013a).

Like generic Big Data, the Upstream Data is also characterized by the 6V:

| | Exploration | Resevoir Engineering & Development | Drilling and Completion | Production |
|---|---|---|---|---|
| **Volume** | Seismic acquisition<br>SEGD | Facilities<br>Reservoir Engineering | Sensors :<br>Flow<br>Pressure<br>ROP | SCADA Sensors :<br>Flow<br>Pressure |
| **Variety** | Structured data :<br>• SEGD<br>• Pre-stack<br>• Post-stack<br><br>Semi-structured :<br>• Implantation Report … | Structured data :<br>• WITSML (XML)<br>• PRODML<br>• RESML<br><br>Unstructured data :<br>• Log Curves / Drilling & Test / Lithology /Cores … | Structured :<br>• WITSML<br><br>Semi-structured :<br>• Final Well Report,<br>• Daily Drilling Report<br><br>Unstructured :<br>• Drilling Log / Gas Log .. etc. | Structured Production data :<br>• PRODML<br>• RESML<br><br>Semi-structured :<br>• Crude Analysis Report |
| **Velocity** | Real Time Data Acquisition :<br>Wide azimuth data acquisition | | Real Time Data Acquisition :<br>Mud Logging / LWD / MWD | Real Time Data Acquisition :<br>SCADA Sensors |
| **Veracity** | Seismic processing | Reservoir Modeling | Sensor calibration | Sensor calibration |
| **Variability** | Seismic Interpretation<br>Geology Interpretation | Reservoir Simulation<br>Combination of seismic, drilling and production data | Data Interpretation & Optimisation | Data Interpretaion |
| **Value** | Navigation,<br>Visualization & Discovery<br>Run integrated asset models | Improve Drilling Program<br>Drive innovation with unconventional resources (shale gas, tight oil) | Reduce costs<br>Reduce Non Productive Time (NPT)<br>Reduce risks<br>Improve HSE performances | Increase speed to first oil<br>Enhancing production |

Fig. 3. Upstream Big Data

## 2. Big Data Still in experimental stage in Oil & Gas Industry

Work on the application of Big Data and Analytics in the Oil & Gas industry is in the experimental stage (Feblowitz, 2012). Only a handful companies have adopted Big Data in the field (Nicholson, 2013):

– Chevron proof-of-concept using Hadoop (IBM BigInsights) for seismic data processing ;
– Shell piloting Hadoop in Amazon Virtual Private Cloud (Amazon VPC) for seismic sensor data ;
– Cloudera Seismic Hadoop project combining Seismic Unix with Apache Hadoop ;
– PointCross Seismic Data Server and Drilling Data Server using Hadoop and NoSQL ;
– University of Stavanger data acquisition performance study using Hadoop.

Much of the software innovation that's key to the digitization of big oil is happening at oil service contracting companies, such as Halliburton and Schlumberger, and big IT providers including Microsoft, IBM, Oracle and Open Source Projects.

### 3. Exploration and Development

By combination of Big Data and advanced analytics in Exploration and Development activities, managers and experts can perform strategic and operational decision-making.

The areas where the analytics tools associated with Big Data can benefit Oil & Gas exploration include:

- **Enhancing exploration efforts:** Historical drilling and production data help geologists and geophysicists verify their assumptions in their analysis of a field where environmental regulations restrict new surveys (Feblowitz, 2012). Combine enterprise data with real-time production data to deliver new insights to operating teams for enhancing exploration efforts (Hems & al, 2013).
- **Assessing new prospects:** Create *competitive intelligence* using Analytics applied to geospatial data, oil and gas reports and other syndicated feeds in order to bid for new prospects (Hems & al, 2013).
- **Identifying seismic traces:** Using advanced analytics based on Hadoop and distributed Database for storage, quick visualization and comprehensive processing and imaging of seismic data (Seshadri, 2013) to identify potentially productive seismic trace signatures previously (Hems & al., 2013).
- **Build new scientific models:** By using high performance computing and based on combination between "historical data" and "Real Time Data Acquisition": Petabyte seismic data sets, Mud Logging, MWD, LWD, Testing, Gamma Ray (Nicholson, 2013) & (Seshadri, 2013).

### 4. Drilling and Completion

Making the drilling platforms and pipeline infrastructure smart to anticipating issues and acting to prevent failures and increase productivity (Seshadri, 2013). In this case, Big Data is used to identify conditions or anomalies that would impact on drilling operations can save human lives and equipments. Real-time information returned from supervisory control and data acquisition systems on well-heads can be used to grasp opportunities that maximize asset performance and optimize production (Hems & al, 2013).

Related areas where analytics can improve drilling and completion operations:

- **Build and assessment of drilling models:** based on all existing well data. These models are incorporated with geologic measurement into drilling processes, such as shale development (Hems &al., 2013). This will refresh models based on incoming sensor data from drill rig and help to optimize drill parameters.
- **Improve Drill Accuracy and Safety:** by early identifying anomalies that would impact drilling and prevent undesired events: kicks, blowout, etc.
- **Drilling optimization:** predictive Analytics help to reduce NPT, by early identifying the negative impacting factors of drilling operations.
- **Cost optimization:** by using scalable compute technologies to determine optimum cost.
- **Real-time decision-making:** must act in real time on drilling data and formation evaluation data and use this data for predictive modeling to facilitate real-time decision-making.
- **Predictive maintenance:** predict drill maintenance/downtime.

### 5. Production

Big Data is of great interest to production and operation work. Being able to predict future performance based on historical results, or to identify sub-par production zones, can be used to shift assets to more productive areas. Oil recovery rates can be improved, as well, by integrating and analyzing seismic, drilling, and production data to provide self-service business intelligence to reservoir engineers.

- **Enhanced oil recovery:** Enhancing oil recovery from existing wells is a key objective for oil and gas companies. Analytics applied to a variety of Big Data at once – seismic, drilling, and production data – could help reservoir engineers map changes in the reservoir over time and provide decision support to production engineers for making changes in lifting methods. This type of approach could also be used to guide fracking in shale gas plays (Feblowitz, 2012).
- **Performance forecasting:** Forecast production at thousands of wells. Aging wells where the forecast does not meet a predetermined production threshold are flagged for immediate remediation (Feblowitz, 2012).
- **Real-time production optimization:** Real-time SCADA and process control systems combined with analytics tools help Oil & Gas producer to optimize resource allocation and prices by using scalable compute technologies to determine optimum commodity pricing. They also, help to make more real time decisions with fewer engineers (Hollingsworth, 2013).
- **Improve Safety and prevent risks:** by early detecting well problems before they become serious – slugging, WAG gas breakthrough.

### 6. Equipment maintenance:

Predictive maintenance is not a new concept for the oil and gas industry, although if you ask a maintenance executive, it does not get the attention and budget it deserves. In upstream, if pressure, volume, and temperature can be collected and analyzed together and compared with the past history of equipment failure, advanced analytics can be applied to predict potential failures. Additionally, many upstream operations are in remote locations or on ships, so being able to plan maintenance on critical assets is important, especially if work requires purchase of specialized equipment (Feblowitz, 2012).

Technicians often use data collected from pumps and wells to adjust repair schedules and prevent or anticipate failure. Better predictive maintenance also becomes possible (Hems & al., 2013):

- **Preventing down time:** Understand how maintenance intervals are affected by variables such as pressure, temperature, volume, shock, and vibration to prevent failure and associated downtime.
- **Optimizing field scheduling:** Use this insight to predict equipment failures and enable teams to more effectively schedule equipment maintenance in the field.
- **Improving shop floor maintenance planning:** Integrate well and tool maintenance data with supply chain information to optimize scheduling of shop floor maintenance.

### 7. Reservoir Engineering

Oil & Gas companies improve understanding of future strategy based on available oil for a better identification of reservoirs and reserves by integrate real-time data into the earth model both on the rig and in the office.

Also, they predict the chances of success of turning reservoir into a production well by:

- **Improving engineering studies:** engage sophisticated subsurface models and conduct detailed engineering studies on wells to identify commercial prospects earlier and with less risk (Feblowitz, 2012).
- **Optimizing subsurface understanding:** Use Big Data tools to understand the earth's subsurface better and to deliver more affordable energy, safely and sustainably.
- **Experiences and learned lessons from drilling operations:** such determination of drilling coordinates through oil shale to optimize the number of wellheads needed for efficient extraction of oil, optimization of drilling resources by not over drilling well site, reducing waste of drilling exploration wells, etc.

### 8. Research & Development

Oil & Gas companies should take advantage of their Research & Development Centers (CRD) or through close collaboration with existing independent centers to explore the potential of Big Data to resolve problems and technical difficulties they encounter in daily operations and capitalize knowledge to improve performance. They must in this case, taking into account the requirements of a shared collaborative environment that supports the flow of total production of the working groups.

To solve problems, R&D workgroups should have permanent eye on the wealth of data contained in the patent databases to be inspired particularly as these databases are considered as Open Big Data.

Patents are a unique and invaluable source for R&D because the information they contain, are generally not published elsewhere. In addition, they have a limited duration after which they can be operated without rights. They can also be fallen (for non-payment of annuities, for example) and may in this case be used freely. Finally, they may only be published in certain countries and thus they are not extended to other, hence the possibility to freely operate in these countries (Baaziz & Quoniam, 2013b).

The automatic patent analysis involves using information contained in a patent by analytics software in order to present outcomes by facilitating up the work of experts to understand developments, actors and subjects appearing in the research (Baaziz & Quoniam, 2013b).

It thus shows the correlations needed to answer common questions (Baaziz & Quoniam, 2013b):

- Importance of the subject and its evolution over time,
- Different technologies and applications involved,
- Who does what? How? (automatic benchmarking of companies),
- What are the trends in research and applications by applicants, inventors, inventors group or country,
- Etc.

Understand the importance of patent information and its proper use at the right time to reduce risks related to operational activities in uncertain environments (Baaziz & Quoniam, 2003b) like developing strategy to address critical design issues or analyzing root causes of design flaw in engineering projects. (Seshadri, 2013).

### 9. Data Management

Big Data and analytics include infrastructure, data organization and management, analytics and discovery for decision support. Infrastructure includes the use of industry-standard servers, networks, storage, and clustering software used for scale-out deployment of Big Data technology. Data organization and management refers to software that processes and prepares all types of data for analysis (Feblowitz, 2012).

### 10. Security

Oil & Gas companies anticipate IT security breaches by using predictive analytics and bolstering security with data from the global protection systems including video monitoring, access control et anti-intrusion.

Also, there is particular interest in deploying complex event processing (CEP) technology to monitor security concerns in the Oil & Gas industry in real time by (Hems & al., 2013):

– Combining data from multiple sources to infer events or patterns that indicate a current or imminent threat.
– Making faster decisions, supported by faster delivery of decision support information, to identify possible threats.
– Predict/prevent cyber-terror acts.

### 11. Health, Safety & Environment

Big Data contributes significantly to reduce risk and optimize costs related to operations and Health, Safety and Environment (Baaziz & Quoniam, 2013b):

– Prevent undesired events while drilling like kicks,
– Predict drill maintenance/downtime, optimize drill parameters and prevent blowout accidents,
– Using weather or workforce scheduling data to avoid creating dangerous conditions for workers and mitigating environmental risks.

## IV. Conclusion

Leading Oil & Gas companies are already began projects to deploying Big Data technologies that can help them track new business opportunities, reduce costs and reorganize operations.

By recognizing the value of the unexploited data assets in supporting fact-based decision-making, Oil & Gas companies establish real cases of Big Data uses. Then they can create enhanced business value based on innovation and able to lead towards a sustainable competitive advantage.

Oil & Gas companies must first proceed to a gap analysis to determine the major requirements of technology and data-management expert staff. This allows a focused investment in mature and proven technologies as well as those who will face the exponential growth of data volumes.

Oil & Gas companies must create new strategies that will help them manipulate these data and use them to support experts in their business process and managers in their decision-making process.